\documentclass[10pt, conference]{IEEEtran}
\usepackage{amsmath,amssymb,amsfonts}
\usepackage{cite}
\usepackage{epsfig}
\usepackage{url}
\usepackage{moreverb}
\usepackage{multirow}
\usepackage{multicol}
\usepackage{balance}
\usepackage{graphicx}
\usepackage{wrapfig}
\usepackage{tcolorbox}
\usepackage[shortlabels]{enumitem}
\usepackage{color}
\usepackage{listings, xcolor}
\usepackage{caption}
\usepackage{algorithmic}
\usepackage[english]{babel}
\usepackage[linesnumbered,ruled,vlined]{algorithm2e}
\usepackage{array}
\usepackage{float}

\usepackage[]{mdframed}

\newcommand{\tcirc}[1]{\textcircled{\raisebox{-0.7pt}{#1}}} 
\newcommand{\tcircA}[1]{\textcircled{\raisebox{-0.7pt}{\scalebox{0.8}{#1}}}} 

\floatstyle{boxed}
\newfloat{boxedFigure}{!ht}{lop}
\floatname{boxedFigure}{Example}
\makeatletter
\newcommand\noVSpace{\@minipagetrue}
\makeatother

\definecolor{dkgreen}{rgb}{0,0.6,0}
\definecolor{gray}{rgb}{0.5,0.5,0.5}
\definecolor{mauve}{rgb}{0.58,0,0.82}

\begin{document}

\title{A Blockchain-Enabled Framework for Storage and Retrieval of Social Data}
\author{
{\rm
 Aishwarya Parab{$^{\dagger}$}, Prakhar Pradhan{$^{\dagger}$},
 Yogesh Simmhan{$^{*}$},
 Arnab K. Paul{$^{\dagger}$}
 
 }\\
{\it {$^{\dagger}$}DaSHLAB - BITS Pilani, KK Birla Goa Campus, India {$^{*}$}Indian Institute of Science (IISc), Bangalore}
\\
\small {$^{\dagger}$}\{\itshape p20220010, f20220992, arnabp\}@goa.bits-pilani.ac.in, {$^{*}$}simmhan@iisc.ac.in
}
\maketitle
\thispagestyle{empty}
\pagestyle{plain}
\label{sec:abstract}

\begin{abstract}
The increasing availability of data from diverse sources, including trusted entities such as governments, as well as untrusted crowd-sourced contributors, demands a secure and trustworthy environment for storage and retrieval. Blockchain, as a distributed and immutable ledger, offers a promising solution to address these challenges. This short paper studies the feasibility of a blockchain-based framework for secure data storage and retrieval across trusted and untrusted sources, focusing on provenance, storage mechanisms, and smart contract security. Through initial experiments using Hyper Ledger Fabric (HLF), we evaluate the storage efficiency, scalability, and feasibility of the proposed approach. This study serves as a motivation for future research to develop a comprehensive blockchain-based storage and retrieval framework.
\end{abstract}

\section{introduction}
\label{sec:intro}

The exponential growth of data from various sources, such as traffic camera networks, road infrastructure sensors, and crowdsourced pollution platforms, has introduced significant challenges in ensuring trust, transparency, and secure access to datasets~\cite{sharifi2024smart}. Citizen services in smart cities, such as Intelligent Transportation Systems (ITS), rely on data from multiple stakeholders, including law enforcement, urban planners, and emergency responders, to improve traffic management, enforce regulations, and optimize urban mobility. However, the integration of data from sources with heterogeneous trust levels, from city institutions to user-generated content, raises concerns about data authenticity, security, and accessibility.

Existing data management systems lack the ability to handle data from dynamic, multi-stakeholder environments with both trusted and untrusted sources~\cite{ahanger2024managing}. Blockchain, with its decentralized and immutable distributed ledger, offers a potential solution for trustworthy data storage and secure data sharing~\cite{paik2019analysis}. 
Further, smart contracts enhance automation, enforce policies, and ensure consistency across transactions, making blockchain ideal for a multi-stakeholder data ecosystem.

However, traditional blockchain systems do not inherently support efficient data retrieval, provenance tracking, or trust evaluation. To address these limitations, we propose a framework that builds on Hyperledger Fabric (HLF)\cite{HyperledgerFabric} and the InterPlanetary File System (IPFS)\cite{ipfs} to enhance data accessibility and management. 

%
We incorporate provenance tracking and efficient querying mechanisms,
focusing on trust management and ensuring accessibility for various users.
Specifically:
\begin{enumerate}[leftmargin=*]
\item We design a blockchain-based architecture for multi-source data integration, ensuring tamper resistance, transparency, and traceability in traffic monitoring applications (\S~\ref{sec:design}).

\item We implement a validation mechanism that incorporates a trust score to assess the reliability of data from both trusted and untrusted sources, leveraging an existing BFT-based consensus algorithm (\S~\ref{sec:design:valid}).

\item We demonstrate efficient data retrieval from the system by combining on-chain and off-chain storage techniques, balancing cost and accessibility (\S~\ref{sec:design:chain}).

\item We evaluate the system on the time required for data storage and retrieval from the blockchain, demonstrating that the additional overhead introduced is minimal (\S~\ref{sec:eval}).
\end{enumerate}

\section{background and Related Work}
\label{sec: related}

Conventional data systems often depend on centralized databases maintained by organizations. 
These systems handle structured data from well-defined sources but struggle with dynamic, heterogeneous data from decentralized environments that are constantly changing~\cite{anagnostopoulos2016handling}. In addition, centralized databases are susceptible to data manipulation, single points of failure, and scalability issues. Moreover, they provide limited support for real-time provenance tracking, making them inadequate for scenarios where data integrity and transparency
are critical in multi-stakeholder systems.

There is much research on using blockchain technology to securely store data and eliminate the need for centralized third-party controllers~\cite{kumi2022blockchain,xie2019survey}. 
Blockchain has been used in IoT and agricultural supply chain management for decentralized access control, using smart contracts to enforce secure data sharing and transparent audit trails~\cite{ayoade2018decentralized,yang2020data}.
Others have examined customized storage structures and off-chain indexing solutions to overcome blockchain's overheads for transaction retrieval and indexing~\cite{wai2019storage}.
Yan et al.~\cite{yan2021handling} have combined the tamper-resistance of blockchain with the quick query processing capabilities of distributed databases. These initiatives highlight how data integrity, scalability, and security problems in different domains can be resolved using blockchain-based storage systems.
However, many of these solutions lack structured data management and efficient queries, making them unsuitable for smart city applications. We address these gaps.

HLF and other hybrid blockchain platforms 
offer a modular architecture, high throughput, and a permissioned environment with adjustable access control. HLF is well-suited when many stakeholders need selective access to sensitive information. Unlike conventional public blockchains, it gives participating organizations control over data accessibility. IPFS, on the other hand, is a peer-to-peer distributed file system enabling decentralized storage and retrieval by storing raw data off-chain and metadata on-chain, thus reducing blockchain storage overhead while ensuring data integrity.


Given the need for a secure, efficient, and scalable solution for managing heterogeneous data from trusted and untrusted sources, we leverage HLF as our base blockchain platform,
and integrate provenance tracking, smart contract automation, and efficient storage mechanisms over it. This can overcome the limitations of existing systems and provide a robust foundation for trusted data management in smart city applications.

\section{System Design}
\label{sec:design}

\begin{figure}[t]
    \centering
    \includegraphics[width=1.0\linewidth]{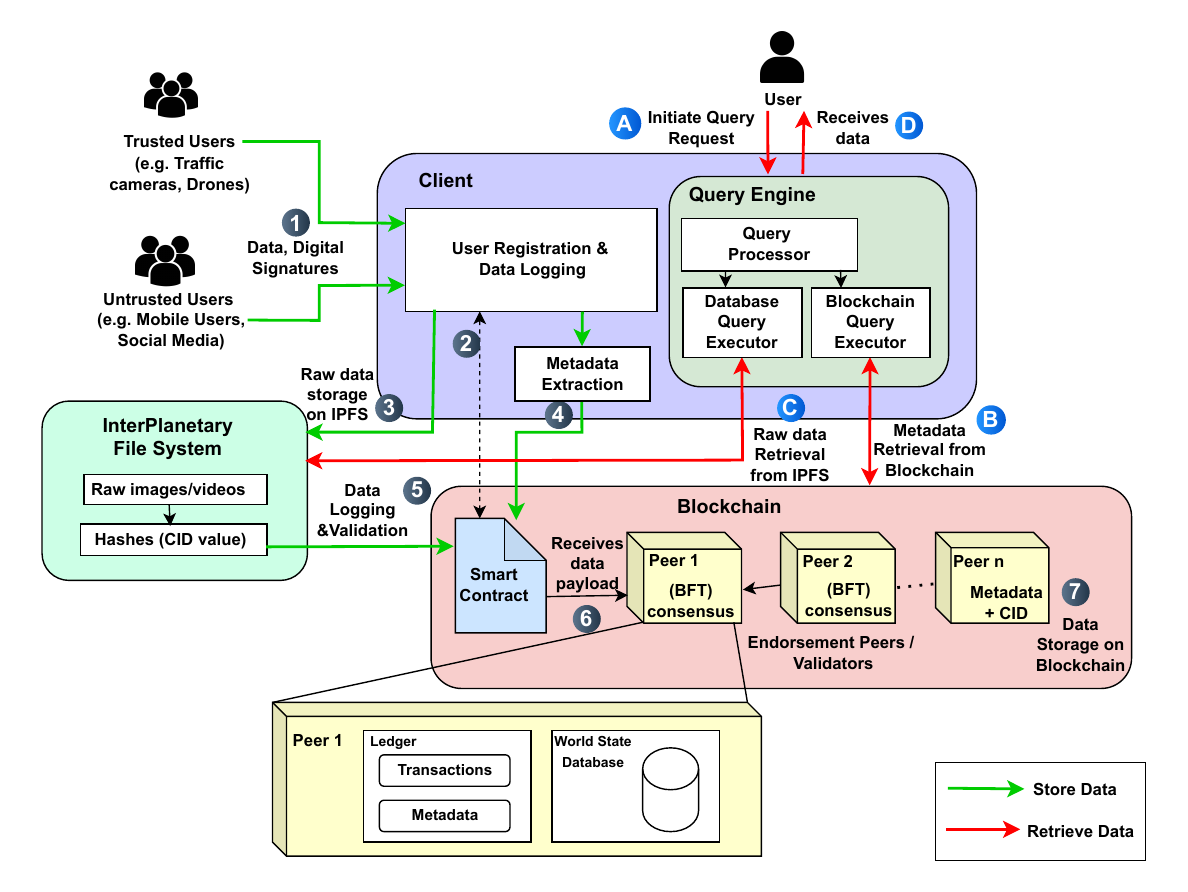}
    \caption{Overview of the system architecture.}
    \label{fig:framework}
\end{figure}

Figure~\ref{fig:framework} outlines our architecture 
that combines IPFS for efficient and secure data storage with HLF for metadata and provenance storage, leveraging smart contracts and Byzantine Fault Tolerance (BFT) consensus to validate and secure data.


\paragraph{Storing data and metadata}
It integrates both trusted users, such as traffic cameras and drones, and untrusted users, such as mobile users and social media platforms. They submit \textit{data} and \textit{digital signatures} to a client \tcirc{1}. The client interacts with a \textit{smart contract} implemented as chaincode, to validate and log the data into the system \tcirc{2}.
The smart contract provides the necessary permissions for the users to submit the transactions to the network. It also verifies if the user's past data contributions align with the blockchain's current records and assesses the digital signatures attached to the data. If discrepancies are detected, the data may require further validation from multiple trusted sources before it is recorded.

Registered and validated users can store data in IPFS \tcirc{3}. Each data entry in IPFS is assigned a unique \textit{cryptographic identifier (CID)} for retrieval. The client extracts metadata from validated data before storing it on the blockchain along with the CID value \tcirc{4}. 

The blockchain network consists of multiple endorsing peers, which function as \textit{validators}. These peers execute smart contracts to verify the submitted data and ensure its integrity before getting added to the ledger \tcirc{5}. The validation process follows a \textit{Byzantine Fault Tolerance (BFT)} consensus mechanism~\cite{tang2024improved}. BFT ensures that the network can achieve agreement on valid transactions even in the presence of malicious peers. Each peer executes the smart contract independently and reaches a consensus before approving a transaction \tcirc{6}. If at least two-thirds of the peers agree on the validity of a transaction, it is considered legitimate and added to the blockchain \tcirc{7}. This approach ensures secure and efficient management of metadata, while the actual data remains in the decentralized IPFS storage.

\paragraph{Retrieving data and metadata}
A \textit{query engine} allows users to retrieve both on-chain metadata and off-chain data. When a user initiates a query request to the client \tcircA{A}, the query processor forwards the request to the appropriate blockchain query executor or database query executor. The blockchain executor retrieves the metadata \tcircA{B} and transaction records from the ledger, while the database executor fetches raw images and videos from IPFS using their CID value \tcircA{C}. This guarantees data integrity by enabling the verification of retrieved data against its metadata stored on the blockchain. Finally, the client provides the requested data to the user \tcircA{D}.

This architecture supports smart city applications like traffic enforcement by capturing road conditions, vehicle movements, and violations via drones, surveillance cameras, and crowdsourced mobiles. Hashed raw data is stored on IPFS, while metadata (e.g., timestamps, locations, vehicle types, violations) is recorded on the blockchain using smart contracts. Law enforcement and analysts query metadata from the blockchain, verifying it against hashed IPFS data, while trust scores assess untrusted sources based on reliability and cross-validation, flagging discrepancies for credibility. This layered approach ensures secure logging, validation, and transparent traffic management.

\subsection{Validators in Blockchain}
\label{sec:design:valid}
A subset of peers in the blockchain act as validators, executing the BFT consensus algorithm to ensure only valid transactions are added to the blockchain. Each validator independently runs the \textit{validation smart contract}, which performs two key checks: (1) Source authentication, verifying the metadata to ensure the data's origin, and (2) Schema verification, ensuring completeness, correct data types, and cryptographic hash integrity.
This HLF smart contract below shows this:
\begin{lstlisting}
    async validateTransaction(ctx, transactionId, metadata, dataPayload) {
    // Source Authentication
    const sourceIsValid = await this.validateSource(ctx, metadata.source);
    if (!sourceIsValid) {
        throw new Error(`Invalid source for transaction ${transactionId}`);}
    // Schema Verification
    const schemaIsValid = this.verifySchema(dataPayload);
    if (!schemaIsValid) {
        throw new Error(`Invalid schema for transaction ${transactionId}`);}
\end{lstlisting}

Validators then vote on the transaction’s validity and share their decisions via a secure peer-to-peer protocol. A transaction is accepted if at least two-thirds of validators reach a consensus. For untrusted sources, validators compute a trust score based on historical accuracy and peer endorsements, storing it on-chain for future reference. Trust measures include historical reliability and cross-validation with trusted data, as they are practical and efficient. Historical reliability predicts trustworthiness by tracking data correctness over time, while cross-validation ensures new inputs match verified information—both with lower computational costs than machine learning-based methods. The BFT mechanism allows the network to tolerate up to one-third of malicious validators. Validators that repeatedly act against the consensus rules (e.g., by endorsing invalid transactions) are flagged and removed from the validator pool.

\subsection{Features of a Chaincode} 
\label{sec:design:chain}
\textit{Chaincodes} are 
smart contract programs in Hyperledger Fabric that define the business logic governing transactions, offering key functionalities in our system.

\paragraph{Role management} The \textit{Admin Enrollment} chaincode administers users. It assigns unique admin IDs with specific permissions, ensuring only authorized personnel can perform administrative actions. It prevents duplication by checking for existing admin IDs before enrollment.
This securely stores admin metadata on the blockchain for verification and auditing. 
\begin{lstlisting}
async enrollAdmin(ctx, adminId) {
    const exists = await this.adminExists(ctx, adminId);
    if (exists) {
        throw new Error(`Admin ${adminId} already exists`);}
    const admin = { role: 'admin', createdAt: new Date().toISOString() };
    await ctx.stub.putState(adminId, Buffer.from(JSON.stringify(admin)));
    return `Admin ${adminId} enrolled successfully`;}
\end{lstlisting}
The \textit{User Registration} chaincode registers users by validating and recording their credentials for audits and accountability.

\paragraph{Data Storage and Retrieval} We use IPFS for efficient data storage, with only data CIDs and metadata stored on-chain to minimize storage costs while preserving data integrity.

The \textit{Data Upload} chaincode uploads data to IPFS, retrieves its CID, and stores it on the blockchain along with metadata. This reduces the cost of blockchain storage and also ensures that data can be efficiently retrieved using the on-chain CID.
\begin{lstlisting}
    async addDataToIPFS(ctx, data) {
    const cid = await ipfsClient.add(data); // Call IPFS client to get CID
    const metadata = { cid, createdAt: new Date().toISOString() };
    const txId = ctx.stub.getTxID();
    await ctx.stub.putState(txId, Buffer.from(JSON.stringify(metadata)));
    return cid;}
\end{lstlisting}

The \textit{Data Retrieval} chaincode retrieves metadata from the blockchain and fetches the corresponding data from IPFS. This ensures efficient and secure data retrieval while maintaining provenance. 


\begin{lstlisting}
    async getDataFromIPFS(ctx, txId) {
    const metadataBytes = await ctx.stub.getState(txId);
    if (!metadataBytes || metadataBytes.length === 0) {
        throw new Error(`No metadata found for transaction ID ${txId}`);   }
    const metadata = JSON.parse(metadataBytes.toString());
    const data = await ipfsClient.get(metadata.cid); // Fetch data from IPFS
    return data;}
\end{lstlisting}

\paragraph{Data provenance} This is a key feature of our system, ensuring trustworthiness, traceability, and integrity. The chaincode uses cryptographic hashes to verify data integrity, preventing tampering and maintaining an immutable record of changes. Metadata stored alongside the CID enables verification of the data’s origin, timestamp, and source, making it essential for applications requiring high levels of trust and compliance.

\section{preliminary evaluation}
\label{sec:eval}

To validate the initial feasibility of our approach, we attempt to answer the following questions:
\begin{itemize}
    \item How does the system perform in terms of time efficiency for storing data on hybrid data storage?
    \item Is the proposed framework scalable for varying data sizes without significant performance degradation?
\end{itemize}

\paragraph{Experimental Setup} We conduct experiments on a private Hyperledger Fabric (HLF) network with one channel, two peer nodes, an orderer node (Docker-deployed), and two IPFS nodes for decentralized storage. Tests were performed on a system with an Intel Core i7 12th Gen processor, A4500 GPU, and 128GB RAM, using Grafana and Hyperledger Explorer for performance monitoring.

\paragraph{Dataset} Our dataset includes 52 traffic videos from static cameras across Bangalore, sourced from the India Urban Data Exchange (IUDX). 
We use the YOLO model to extract video frames, identifying and classifying vehicles (e.g., cars, trucks, two-wheelers) along with metadata like timestamps, colors, and location coordinates. Figure~\ref{fig:metadataexample} illustrates an example of the extracted metadata record.

\begin{figure}[h!]
\centering
\begin{tcolorbox}[colback=gray!4, boxrule=0.2mm, boxsep=1mm, left=0.2mm, right=0mm, top=0mm, bottom=0mm]
\scriptsize
\texttt{metadata \{\\
    \ \ "label": "truck",\\
    \ \ "confidence": 0.41042160987854004,\\
    \ \ "bounding\_box":\{ "x1":755, "y1":82, "x2":1023, "y2":506 \},\\
    \ \ "timestamp": "2024-07-10T05:55:46.304199Z",\\
    \ \ "color": "yellow",\\
    \ \ "location": \{ "latitude": 40.712303728004414, "longitude": -74.00629823104597 \}\\
\}}
\end{tcolorbox}
\caption{Sample metadata record extracted from the images.}
\label{fig:metadataexample}
\end{figure}

\begin{figure}[!ht]
\centering
\begin{minipage}[c]{0.47\linewidth}
\centering
\includegraphics[width=0.99\linewidth]{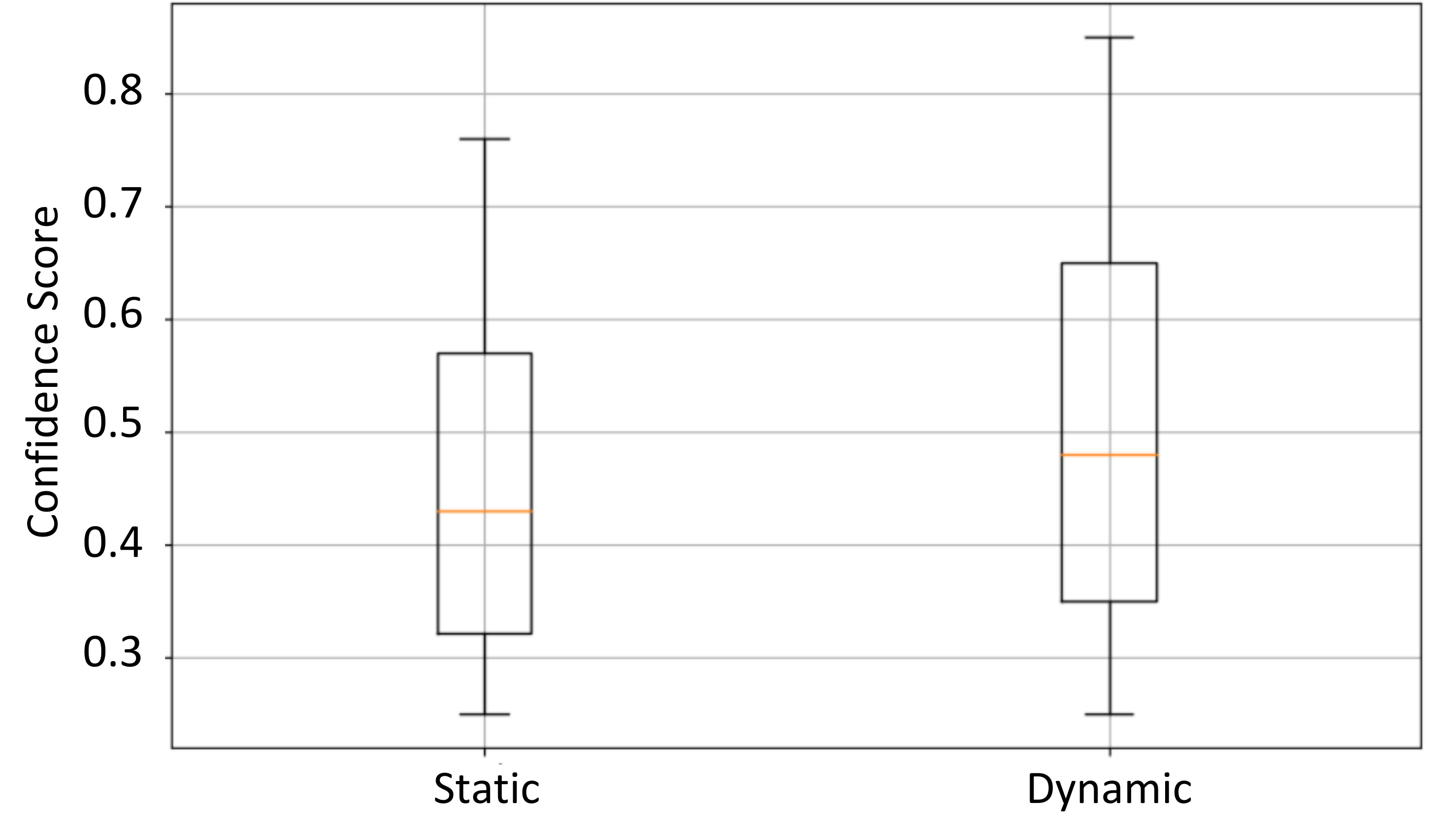}
\caption{Confidence scores for static and drone-captured data.}
\label{fig:confidencescore}
\end{minipage}
\hfill
\begin{minipage}[c]{0.47\linewidth}
\centering
\includegraphics[width=0.99\linewidth]{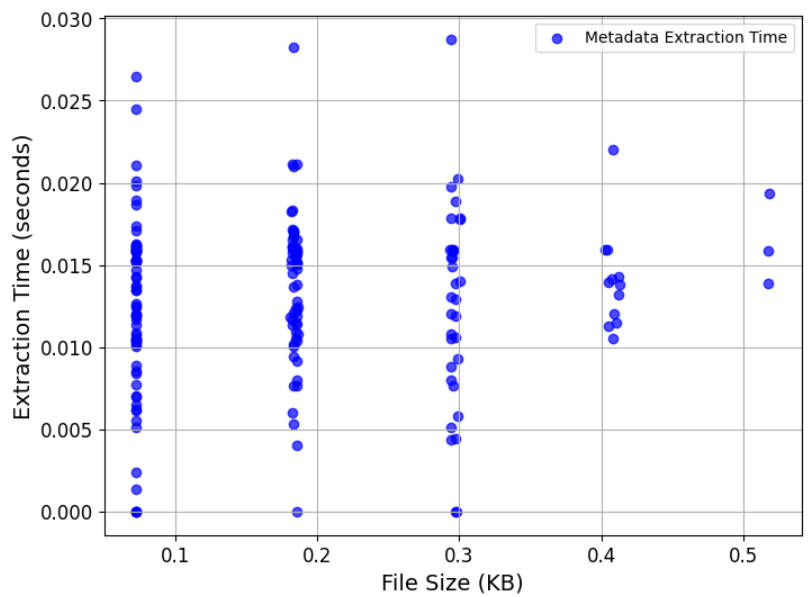}
\caption{Metadata extraction time from raw traffic images, varying by file size.}
\label{fig:metadataextraction}
\end{minipage}
\end{figure}

\paragraph{Results and Analysis}
The observed metrics reflect system design choices that maintain data integrity with minimal computational overhead through provenance tracking and efficient metadata handling. Storing metadata on-chain and raw data on IPFS enables quick retrieval while reducing blockchain storage costs. To assess object detection consistency, we compared frames from static cameras and drone-captured datasets. As shown in Figure~\ref{fig:confidencescore}, static cameras yielded higher and more stable confidence scores due to consistent capture conditions, while drone data showed greater variability from motion blur, altitude changes, and environmental factors. Addressing this variability is vital for robust drone-based traffic monitoring.

Our preliminary evaluation shows the framework’s feasibility and efficiency, while future work will focus on large-scale validation with diverse datasets, assessing scalability and fault tolerance under various blockchain configurations, and enhancing trust scoring with advanced techniques like multi-source consensus and anomaly detection.

The scatter plot shown in Figure~\ref{fig:metadataextraction} shows the time taken to extract metadata from frames of various sizes. Most data points are clustered around smaller file sizes (under 0.5 KB), with extraction times typically ranging from 0.002 to 0.01 seconds. Contrary to the initial assumption, several smaller file sizes still exhibit relatively longer extraction times, suggesting that the time taken is not strictly linear with file size. This variance could be attributed to differences in metadata complexity, file encoding formats, or the computational overhead associated with processing certain data structures. Generally, smaller file sizes tend to result in shorter metadata extraction times, but outliers indicate that other factors may also play a role.

\begin{figure}[!ht]
\centering
\begin{minipage}[c]{0.47\linewidth}
\centering
\includegraphics[width=0.95\linewidth]{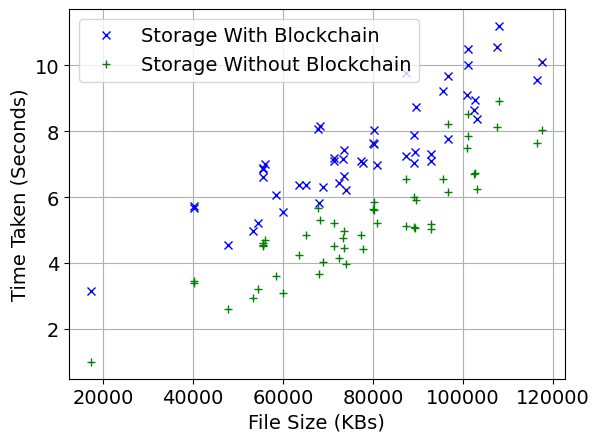}
\caption{\textit{Storage time in IPFS across file sizes}, with and without blockchain overheads.}
\label{fig:datastorage}
\end{minipage}
\hfill
\begin{minipage}[c]{0.47\linewidth}
\centering
\includegraphics[width=0.95\linewidth]{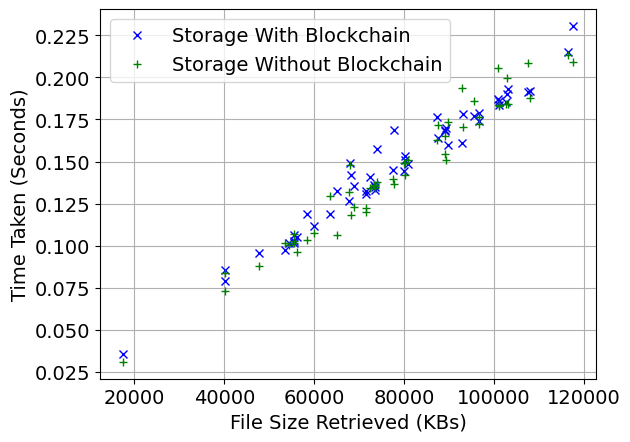}
\caption{\textit{Retrieval time in IPFS across file sizes}, with and without blockchain overheads.}
\label{fig:dataretrieval}
\end{minipage}
\end{figure}

Understanding IPFS scalability is crucial for large datasets, so we evaluate storage time with and without blockchain integration. Figure~\ref{fig:datastorage} compares the time taken to store files of varying sizes on IPFS, with and without blockchain overhead. Results show a nearly linear correlation between file size and storage time in both cases, demonstrating that blockchain integration adds minimal overhead while preserving data integrity and provenance.

Figure~\ref{fig:dataretrieval} illustrates retrieval times, comparing metadata access from the blockchain and data retrieval via CID from IPFS. While retrieval time increases with file size, blockchain overhead remains minimal, ensuring efficiency even for large datasets. Since reading from the blockchain does not incur gas costs, the process remains computationally inexpensive.

These findings confirm the framework’s scalability and adaptability, with further opportunities to optimize drone data preprocessing.

\label{sec:conclusion}

\section{Conclusion}

We propose a blockchain-enabled framework for secure and efficient data storage and retrieval that addresses challenges in integrating trusted and untrusted data sources. Leveraging Hyperledger Fabric and IPFS, our approach ensures data provenance, integrity, and accessibility across stakeholders with minimal overhead in storage and retrieval while maintaining scalability and reliability.  The analysis of static and drone-captured datasets highlights the framework’s adaptability to diverse data sources, proving its utility in real-world applications like traffic management and urban planning. Future work will involve testing with larger, real-world datasets, exploring additional monitoring scenarios, integrating advanced trust scoring, and evaluating performance under different blockchain configurations.

\bibliographystyle{unsrt}
\bibliography{refs}
\end{document}